\documentclass[twocolumn, 
  prb,preprintnumbers,amsmath,amssymb,floatfix]{revtex4}

\usepackage{graphicx}
\usepackage{dcolumn}

\begin{document}

\title{Competing Phases, Strong Electron-Phonon Interaction and Superconductivity in Elemental Calcium under High Pressure}
\author{Z. P. Yin$^1$, F. Gygi$^2$, and W. E. Pickett$^1$}
\affiliation{$^1$Department of Physics, University of California Davis, 
  Davis, CA 95616\\
             $^2$Department of Applied Science,  University of California Davis,Davis, CA 95616}
\date{\today}
\begin{abstract}
The observed ``simple cubic'' (sc) phase of elemental Ca at room temperature in the 32-109 GPa
range is, from linear response calculations, dynamically unstable.
By comparing first principle calculations of the enthalpy for five
sc-related (non-close-packed) structures, we find that all five structures
compete energetically at room temperature in the 40-90 GPa range, and three do so in
the 100-130 GPa range. Some competing structures below 90 GPa are dynamically stable, 
{\it i.e.}, no imaginary frequency, 
suggesting that these sc-derived short-range-order local structures exist 
locally and can account for the observed (average) ``sc'' diffraction
pattern.  In the dynamically stable phases below 90 GPa, some
low frequency phonon modes are present, contributing to strong electron-phonon (EP) coupling as well as arising
from the strong coupling.
Linear response calculations for two of the structures over 120 GPa 
lead to critical temperatures in the 20-25 K range as is observed, and do
so without unusually soft modes.
\end{abstract}
\maketitle

\section{Introduction}
One of the most unanticipated developments in superconducting critical
temperatures (T$_c$) in the past few years has been achievement
of much higher values of T$_c$ in elemental superconductors by the
application of high pressure, and that these impressive superconducting
states evolve from simple metals
(not transition metals) that are non-superconducting at ambient pressure.
The first breakthrough arose in Li, with T$_c$ approaching\cite{Li-exp1,Li-exp2} 20 K,
followed by yttrium\cite{Y-exp,Sc-exp} at megabar pressure also superconducting up to 20 K
and showing no sign of leveling off.
Both of these metals have electron-phonon
(EP) coupled pairing, according to several linear response 
calculations\cite{Li-th1,Li-th2,Y-th1,Y-th2} of the phonon
spectrum, electron-phonon coupling (EPC) strength, and application of Eliashberg theory.
These impressive superconductors have been surpassed by Ca, with T$_c$ as high
as 25 K reported\cite{Ca-exp} near 160 GPa.  Perhaps more unusual is the report, from
room temperature x-ray diffraction (XRD), of a simple cubic (hence far from 
close-packed) structure
over a volume reduction
of 45$\rightarrow$30\% (32-109 GPa).  Whether these two unique phenomena
are connected, and in what way, raises fundamental new issues in an
area long thought to be well understood. 

Face-centered cubic (fcc, Ca-I) at ambient pressure, calcium transforms 
at room temperature to
body-centered cubic (bcc, Ca-II) at\cite{olijnyk} 20 GPa, 
is identified as simple cubic (sc, Ca-III) in the
very wide 32-109 GPa range as mentioned above, and shows additional phases (Ca-IV, Ca-V) at even
higher pressures. 
A sc structure for an element is
rare, occurring at ambient pressure only in polonium and 
under pressure only in a handful of elemental metals.\cite{Po-exp1,Po-exp2}  This
identification of a sc structure for Ca is particularly problematic, 
since it has been shown by
linear response calculations of the phonon spectrum by a few groups\cite{ref1,ref2,bonev}
that (at least at zero temperature) sc Ca is highly unstable dynamically at 
all
volumes (pressures) in the region of interest. Since these calculations
are reliable for such metals, there are basic questions about the ``sc''
structure itself.

\section{Comparison to related metals}
Strontium, which is isovalent with Ca, like Ca superconducts under pressure and 
undergoes a series of structural transitions
from close-packed structure to non-close-packed structure at high pressure.
Sr transforms from a fcc phase to a bcc phase at 3.5 GPa and then transforms to Sr-III at 24 GPa, 
to Sr-IV at 35 GPa and to Sr-V at 46 GPa.\cite{Mizobata} 
The Sr-III structure was first believed to be a distorted sc 
and later found to be an orthorhombic structure.\cite{Phusittrakool} 
However, later experiments have found that there are two phases coexisting in the Sr-III phase, 
namely, a tetragonal phase with a distorted $\beta$-tin structure 
and an unidentified additional phase.\cite{Phusittrakool} 
The Sr-IV structure is very complex and was shown recently to be a monoclinic structure 
with the $Ia$ space group and 12 atoms per unit cell.\cite{Bovornratanaraks} 
The structure is more complex in Sr-V, 
and was identified as an incommensurate structure similar to that of Ba-IV.\cite{McMahon} 
Sr begins to superconduct at 20 GPa, its T$_c$ is 8 K at 58 GPa, 
and is believed to be higher beyond 58 GPa.\cite{Mizobata}

Scandium, with one more ($3d$) electron than Ca, undergoes phase transitions 
from hcp to Sc-II at 20 GPa and to a Sc-III phase at 107 GPa.\cite{Zhao, Debessai} 
Although Sc is conventionally grouped together with Y and 
the lanthanide metals as the rare-earth metals, 
due to their similarities in their outer electron configurations, 
its structural transition sequence is rather different from 
the common sequence of lanthanide metals and Y, which follow the pattern 
hcp$\rightarrow$ Sm-type$\rightarrow$ dhcp$\rightarrow$ 
fcc$\rightarrow$ distorted fcc. 
The Sc-II structure is complex, 
and was recently found to be best fitted to a pseudo bcc structure with 24 atoms in the unit cell.\cite{Zhao} 
The structure of Sc-III is not identified to date. 
Sc begins to superconduct at 20 GPa. Its T$_c$ increases monotonically to 19.6 K with pressure to 107 GPa. 
Its T$_c$ drops dramatically to 8 K at the phase transition from Sc-II to Sc-III around 107 GPa.\cite{Debessai}

Considering the close relation of Sc and Sr to Ca in the periodic table and the similar superconducting
properties under pressure, it could be expected that Ca under pressure 
should have more complex structures, rather than
the observed sc structure. In fact, Olijnyk and Holzapfel\cite{olijnyk} 
observed that their Ca sample transformed from sc to an unidentified complex structure at 42 GPa.

So far the higher pressure phases Ca-IV and Ca-V have attracted the most
attention, and considerable progress has been made in identifying these
phases through a combination of experimental\cite{Ca-exp,Ca-exp2, fuji} and theoretical\cite{yao, Ca-MD-Cmca, Arapan} work.
However, satisfactory agreement between experimental and theoretical work is still lacking.  
Ca-IV is identified as a Pnma space group by Yao {\it et al.}\cite{yao} but P4$_3$2$_1$2 
symmetry by Ishikawa {\it et al.}\cite{Ca-MD-Cmca} and Fujihisa {\it et al.}\cite{fuji}
Ca-V seems clearly to have a Cmca space 
group\cite{Ca-MD-Cmca,fuji,yao}, however, 
the calculated enthalpy in the Pnma structure is much lower
than in other structures (including Cmca structure) at pressures over 140 GPa.
Also in the experimental work of Fujihisa {\it et al.}\cite{fuji}, 
the fitting of their XRD patterns to the anticipated P4$_3$2$_1$2 and Cmca 
space groups were not satisfactory and 
other possibilities still exist.
In the recent work of Arapan, Mao, and Ahuja\cite{Arapan}, an incommensurate 
structure similar to Sr-V and Ba-IV structures
was proposed for Ca-V phase. 
Therefore the nature of the Ca-IV and Ca-V phases is still not 
fully settled.

While helping to forge an understanding the structure of Ca-IV and Ca-V and its impressive superconducting T$_c$ is one goal
of the present work,
our focus has been to understand the enigmatic ``sc'' Ca-III phase
where relatively
high T$_c$ emerges and increases with pressure, a phase that XRD at
room temperature (T$_{R}$) identifies as primitive simple cubic.\cite{Ca-exp2}  In this 
pressure range sc Ca becomes favored over the more closely packed fcc and bcc 
structures, but the dynamical (in)stability was not calculated
by Ahuja {\it et al.}\cite{ahuja}
We report here first principles calculations of
the enthalpy of five crystal structures
(with space groups sc, I${\bar 4}$3m, P4$_3$2$_1$2, Cmca, and Pnma), and linear response
calculations of EPC, that helps to clarify
both the structural and superconducting questions.

\section{Theoretical Approach}

\subsection{Competing Structures}
\begin{figure}[tbp]
{\resizebox{7.8cm}{5.6cm}{\includegraphics{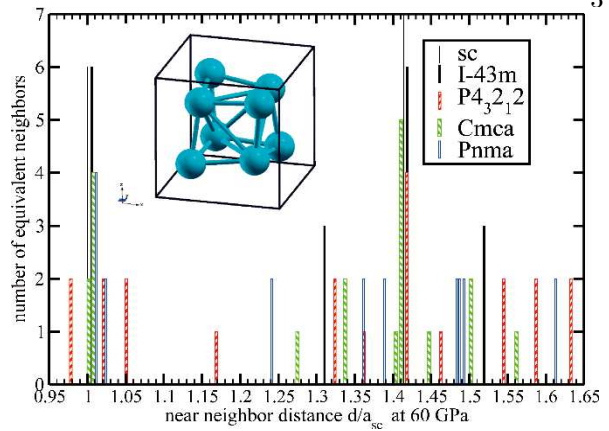}}}
\caption{(color online) Local coordination of the five structures of Ca,
plotted as number of neighbors versus the distance $d$ relative to the
cubic lattice constant $a_{sc}$ with the same density.  The inset
shows the unit cube of the I${\bar 4}$3m structure (which contains two
primitive cells); this structure retains six near neighbors at equal
distances but three different second neighbor distances.
The P4$_3$2$_1$2 and Pnma structures can be regarded to be seven-coordinated,
albeit with one distance that is substantially larger than the other six.
}
\label{nn60}
\end{figure}

The most unstable modes of sc Ca are transverse [001]-polarized zone
boundary modes along the (110) directions.  A linear combination of
the eigenvectors of this mode at different zone boundary points
leads to a body-centered four-atom cell in the
space group I${\bar 4}$3m, whose local coordination is shown
in the cubic cell in the inset of
Fig \ref{nn60}, and has a clear interpretation as a buckled sc lattice.
This structure, when relaxed, has no dynamical instabilities.

The I$\bar{4}$3m structure is just one kind of distortion from the sc structure.
There are many kinds of other possible distortions.
Actually several other structures including Pnma, Cmca and P4$_3$2$_1$2 
were proposed for the high pressure Ca-IV and Ca-V phases.\cite{fuji, yao, Ca-MD-Cmca, Arapan}
Their structural details are listed in Table \ref{Ca-structure-table}  and their structures are pictured in reference [\onlinecite{fuji, yao, Ca-MD-Cmca}].
I$\bar{4}$3m is a body-centered cubic structure,  Pnma and Cmca Ca are orthorhombic,
and P4$_3$2$_1$2 has a tetragonal symmetry.
All are closely related to sc structure. For example, I$\bar{4}$3m turns to simple cubic if $x$=0.25,
and the Cmca structure becomes a sc structure if $a=b=c$ and $y=z=0.25$.

\begin{table}
\caption{Detailed structural data of the I$\bar{4}$3m, $Pnma$, $Cmca$ and $P4_32_12$ Ca. 
(SG: space group; WP: Wyckoff position; AC: atomic coordinates.) }
\label{Ca-structure-table}
\begin{tabular}{|c|c|c|c|c|c|c|}
\hline
SG    & No.   &  WP & AC & $x$ & $y$ & $z$ \\
\hline
I$\bar{4}$3m   & 217   & 8c                & ($x$, $x$, $x$)    & $\sim$ 0.2 & &  \\
$Pnma$         & 62    & 4c                & ($x$, 1/4, $z$)    & $\sim$ 0.3 & & $\sim$ 0.6 \\
$Cmca$         & 64    & 8f                & (0, $y$, $z$)      &      & $\sim$ 0.3 & $\sim$ 0.2 \\
$P4_32_12$     & 96    & 8b                & ($x$, $y$, $z$)    & $\sim$ 0   & $\sim$ 0.3 & $\sim$ 0.3 \\
\hline
\end{tabular}
\end{table}

\subsection{Calculational Methods}
We have used the full potential
local-orbital (FPLO) code,\cite{fplo}
the full-potential linearized augmented plane-wave (FPLAPW) + local orbitals (lo) method as implemented in WIEN2K,\cite{wien2k}, the Qbox
code\cite{qbox} and the PWscf code \cite{pwscf} 
to do various structural optimizations and electronic structure
calculations, and check for consistency among the results.
For the enthalpy calculations we used the PWscf code.\cite{pwscf}
Both Qbox and PWscf
use norm-conserving
pseudopotentials, while the FPLO and WIEN2K codes are all-electron and full potential codes.
The linear-response calculations of phonon spectra and electron-phonon
spectral function $\alpha^2F(\omega)$ were done using the all-electron,
full potential LMTART
code.\cite{lmtart, lmtart1}

The parameters used in PWscf for the structural optimizations and enthalpy calculations were:
wavefunction planewave cutoff energy of 60 Ry, density planewave cutoff energy of 360
Ry, $k$ mesh samplings (respectively, number of irreducible $k$
points) 24*24*24 (455), 32*32*32
(897), 24*24*8 (455), 24*24*24 (3614), 24*32*32 (6562) for sc, I${\bar 4}$3m,
P4$_3$2$_1$2, Cmca, and Pnma structure, respectively.  Increasing the
number of $k$ points lowers the enthalpy by only 1-2 meV/ Ca
almost uniformly for all structures, resulting in negligible change in
volume, lattice constants, and internal coordinates. 
In these calculations, we used a Vanderbilt ultrasoft pseudopotential\cite{USPP} 
with Perdew-Burke-Ernzerhof\cite{PBE} (PBE) exchange correlation functional and nonlinear core-correction,
 which included semicore $3s3p$ states as well as $4s3d$ states in valence states.

\vskip 2mm

\section{The enthalpies}

\begin{figure}[tbp]
{\resizebox{7.8cm}{5.6cm}{\includegraphics{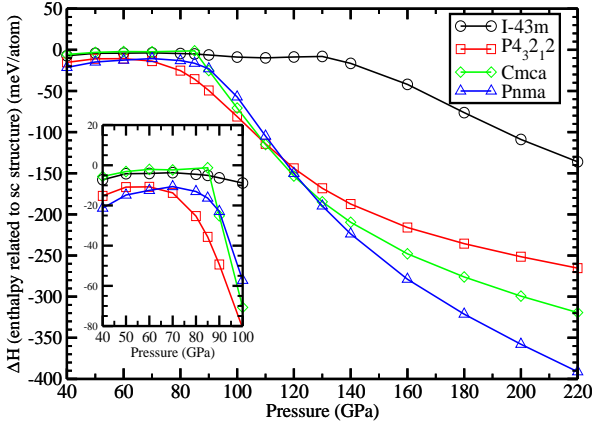}}}
\caption{(color online) Plot of the enthalpy H(P) of the four distorted Ca
structures relative to that for Ca in the simple cubic structure.  The
inset gives an expanded picture of the 40-100 GPa regime.
   }
\label{HofP}
\end{figure}

We have calculated enthalpy H(P) curves for each structure
in the pressure range 40-220 GPa based on density functional methods\cite{HK, KS}  
using the PWscf code.\cite{pwscf} 
Several energy differences and relaxations were checked with the Qbox,\cite{qbox}
FPLO,\cite{fplo} and WIEN2K\cite{wien2k} codes.
In the 40-70 GPa range,
all five of the structures we have studied
have enthalpies that differ by less than 20 meV/Ca (230 K/Ca), as shown in
Fig. \ref{HofP}.  In the 80-100
GPa range, the P4$_3$2$_1$2 phase is marginally the more stable phase.
Three phases are degenerate, again within 20 meV/Ca, in the 100-130 GPa
region and are almost exactly degenerate around 110-115 GPa.
Thus at room temperature all five phases, including the sc one,
are thermodynamically accessible up to 80-90 GPa, above which the
sc and I${\bar 4}$3m structures become inaccessible.  The other three
phases remain thermally accessible to 130 GPa.
Above 140 GPa, the Pnma phase becomes increasingly more stable
than the others.  

Our results agrees well with the results reported recently by
Yao {\it et al.}\cite{yao} and 
Ishikawa {\it et al.}\cite{Ca-MD-Cmca} in their corresponding pressure range.
At low pressure, our result is apparently different from the result by Arapan, Mao, and Ahuja.\cite{Arapan}
In their results, sc Ca has the lowest enthalpy from 40 GPa to 77 GPa, lower than the P4$_3$2$_1$2 and Cmca
structures.  A possible reason is that the authors might not have taken into account the change in shape and internal coordinates
of the Cmca structure in the 70-80 GPa pressure range. In our calculation, b/a=1.0003 and internal coordinates y=0.254, 
z=0.225 at 70 GPa (and similarly below) change dramatically to b/a=1.0594, y=0.349 and z=0.199 at 80 GPa 
(and similarly above).

\vskip 2mm

Although equally dense, quasi-degenerate, and related to the sc structure, 
these structures
differ in important ways from the sc structure and each other.  In Fig. \ref{nn60}
the distribution of (first and second) neighbor distances $d$, relative to
the sc lattice constant $a_{sc}$, are pictured.  The collection of distances cluster
around $d/a_{sc} \sim 0.97-1.05$ and, more broadly, around $\sqrt{2}$.  In an
ensemble of nanocrystallites of these phases, the radial distribution function
in the simplest picture
should look like a broadened version of the sc one.  For Ca the actual microscopic
configuration at room temperature, where fluctuations (spacial and temporal)
can occur among these phases (whose enthalpies differ by less than 
$k_B$T$_{R}$ per atom), 
will no doubt be much more complex.  However, this
simplistic radial distribution plot makes it plausible that the resulting thermal and
spatial distribution of Ca atoms
will produce an XRD pattern more like simple cubic than any other simple
possibility.  Teweldeberhan and Bonev have noted the near degeneracy of some
of these phases in the 40-80 GPa region, and suggest that the T=0 structure 
is Pnma in the 45-90 GPa range\cite{bonev}, which is consistent with our results if
the P4$_3$2$_1$2 structure is not included.

\section{Stability and lattice dynamics}
The structural stability of the (quasi-degenerate) structures we have studied provide 
insight into behavior of Ca under pressure. 
Linear response calculations were performed using the LMTART code\cite{lmtart, lmtart1} to evaluate EPC.   

{\it 60-100 GPa.}
The I${\bar 4}$3m and 
Pnma structures are mostly dynamically stable from 60-100 GPa 
according to our linear response calculations,
but there are very soft zone boundary modes that verge on
instability (small imaginary frequencies) at some pressures. 
The Cmca and P4$_3$2$_1$2 structures are unstable over this entire pressure range; 
note that their structures are close to the sc structure. 
However, they are close to stable with very soft phonons at 100 GPa, where they were distorted
far enough from the sc structure. 

\begin{figure}[tbp]
{\resizebox{7.8cm}{5.6cm}{\includegraphics{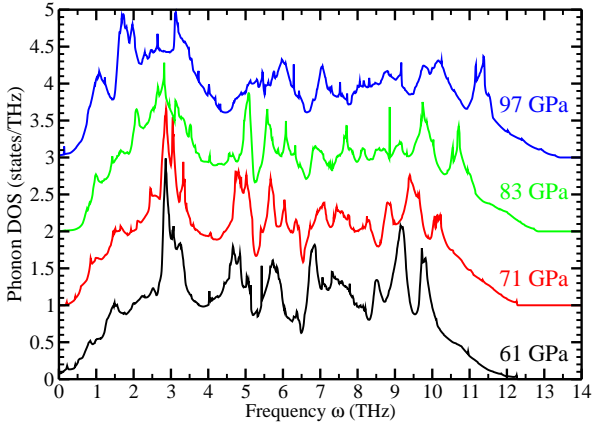}}}
\vskip 10mm
{\resizebox{7.8cm}{5.6cm}{\includegraphics{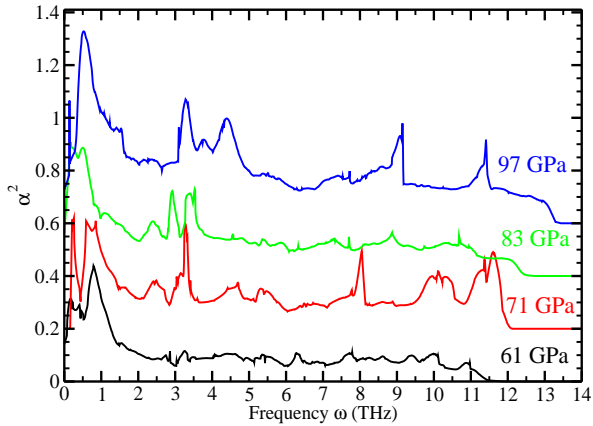}}}
\vskip 10mm
{\resizebox{7.8cm}{5.6cm}{\includegraphics{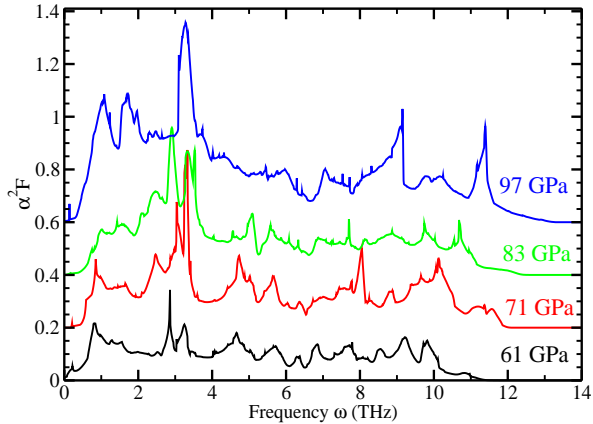}}}
\caption{(color online) Plot of $\alpha$$^2$F($\omega$) (lower panel), 
$\alpha$$^2$($\omega$) (middle panel), 
and phonon DOS (upper panel) of I${\bar 4}$3m structure at about 61, 71, 83 and 97 GPa.
This regime is characterized by strong coupling $\alpha^2(\omega)$ at very low
frequency.}
\label{alpha2F}
\end{figure}

\begin{figure}[tbp]
{\resizebox{7.8cm}{5.6cm}{\includegraphics{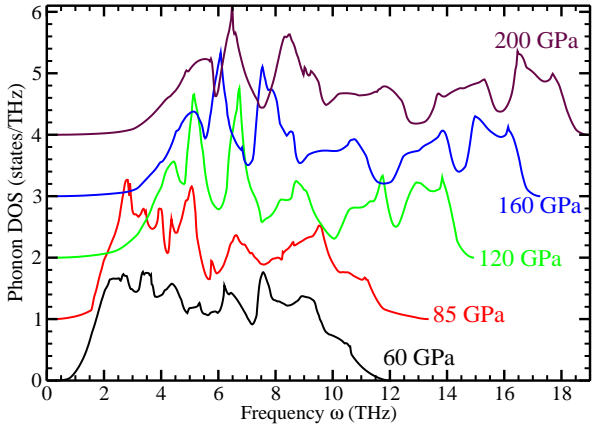}}}
\vskip 10mm
{\resizebox{7.8cm}{5.6cm}{\includegraphics{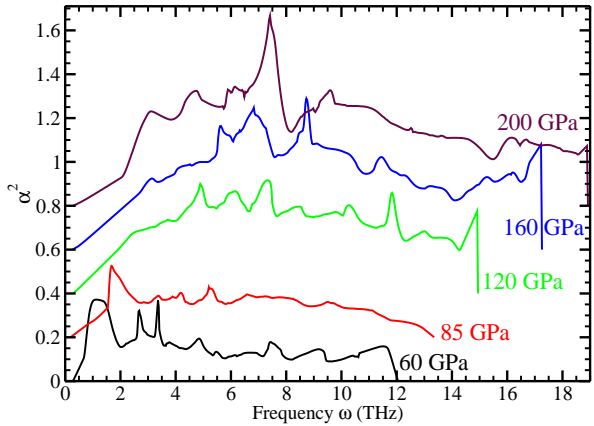}}}
\vskip 10mm
{\resizebox{7.8cm}{5.6cm}{\includegraphics{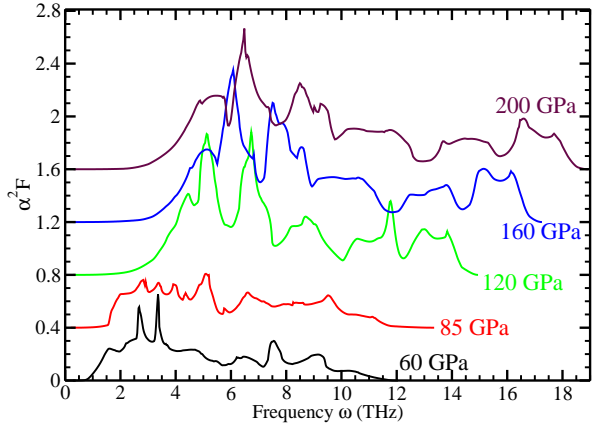}}}
\caption{(color online) Plot of $\alpha$$^2$F($\omega$) (bottom panel), 
$\alpha$$^2$($\omega$) (middle panel),
and phonon DOS (upper panel) of Pnma structure at about 60, 85, 120, 160, and 200 GPa.
The main trends are the stiffening of the modes with increasing pressure, and the
retention of coupling strength $\alpha^2(\omega)$ over a wide frequency range.}
\label{alpha2F-Pnma}
\end{figure}

A rather common feature among these structures in this pressure range 
is softening of modes at the zone boundary,
with associated low frequency weight in the spectral function $\alpha$$^2$($\omega$) that can be
seen in Fig. \ref{alpha2F} and 
Fig. \ref{alpha2F-Pnma}.  Such low frequency weight contributes strongly to
$\lambda$, though the contribution to T$_c$ is better judged\cite{Y-th1} 
by $<\omega>\lambda$ or
even $<\omega^2>\lambda$.
With increase of pressure, the peaks move towards lower frequency, $\lambda$ increases, 
and the structures approach instability.
These results are consistent with the changes of structure parameters we obtain in the process of 
calculating the enthalpies, where all four structures evolve further 
from the sc structure with
increase of pressure. 

{\it Above 100 GPa.}
At the highest pressures studied (by us, and experimentally), the crystal structures
deviate more strongly from the sc structure.
Of the structures we have considered,  the P4$_3$2$_1$2 one becomes favored
and also is structurally stable around 110 GPa.
This stability is consistent with the observed transition from the 
sc structure to the Ca-IV structure
at room temperature.
The dramatic drop in the electrical resistance at around 109 GPa 
is also consistent with a 
transition from a locally disordered phase to a crystalline material.\cite{Ca-exp}

\vskip 2mm

\begin{figure}[tbp]
{\resizebox{7.8cm}{5.6cm}{\includegraphics{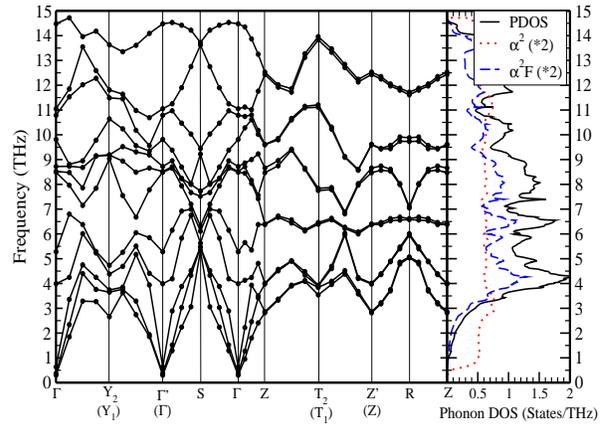}}}
\caption{Phonon spectrum,  phonon DOS, $\alpha^2$ and $\alpha^2$F of Cmca Ca
at 0.251 V$_0$ ($\sim$ 130 GPa from PwSCF).
The high symmetry points are $\Gamma$(0, 0, 0), Y$_1$(1, 0, 0), Y$_2$(0, 1, 0), $\Gamma^{\prime}$(1, 1, 0),
 S(0.5, 0.5, 0), Z(0, 0, 0.5), T$_1$(1, 0, 0.5), T$_2$(0, 1, 0.5), 
Z$^{\prime}$(1, 1, 0.5) and R(0.5, 0.5, 0.5)
in the units of (2$\pi$/a, 2$\pi$/b, 2$\pi$/c).
   }
\label{A2F}
\end{figure}

In the pressure range of 110-140 GPa,  
the P4$_3$2$_1$2, Cmca, and Pnma structures become quasi-degenerate again.
Linear response calculations of the Pnma structure at 120 GPa and above, and 
of the Cmca structure at around 130 GPa indeed
show strong coupling with $\lambda$ $>$ 1.0 in all the cases. 
Unlike what was found below 100 GPa, 
there are no longer very low frequency phonons
(see Fig. \ref{alpha2F-Pnma} and Fig. \ref{A2F}). 
The coupling strength is spread over frequency, peaking for mid-range frequency phonons. 

Another interesting feature arises in the $\alpha$$^2$($\omega$) curves, which reveal
that the coupling matrix elements become relatively uniform
across most of the frequency range (except the uninteresting acoustic modes below 2 THz)
at pressures over 120 GPa in Pnma structure
and at 130 GPa in Cmca structure; this behavior is evident in Fig. \ref{alpha2F-Pnma} 
and especially in Fig. \ref{A2F} where the results for the Cmca structure at 130 GPa
are pictured. 
This characteristic is fundamentally different from that below 100 GPa, discussed above.

At pressures over 140 GPa, the Pnma structure is clearly favored in our 
calculation, and linear response calculations
indicate the structure is dynamically stable. 
The overall results are evident in Fig. \ref{alpha2F-Pnma}, which shows that the structures
remain stable (no imaginary frequencies) and the lattice stiffens smoothly with
increasing pressure, and in Fig. \ref{T$_c$s} that shows that strong electron-phonon coupling persists
and T$_c$ remains high.
In this high pressure range, 
the incommensurate structure proposed by Arapan, Mao, and Ahuja\cite{Arapan} 
at pressure over 130 GPa is also a possibility.
\vskip 2mm
 
\section{Coupling Strength and T$_c$}
\begin{figure}[tbp]
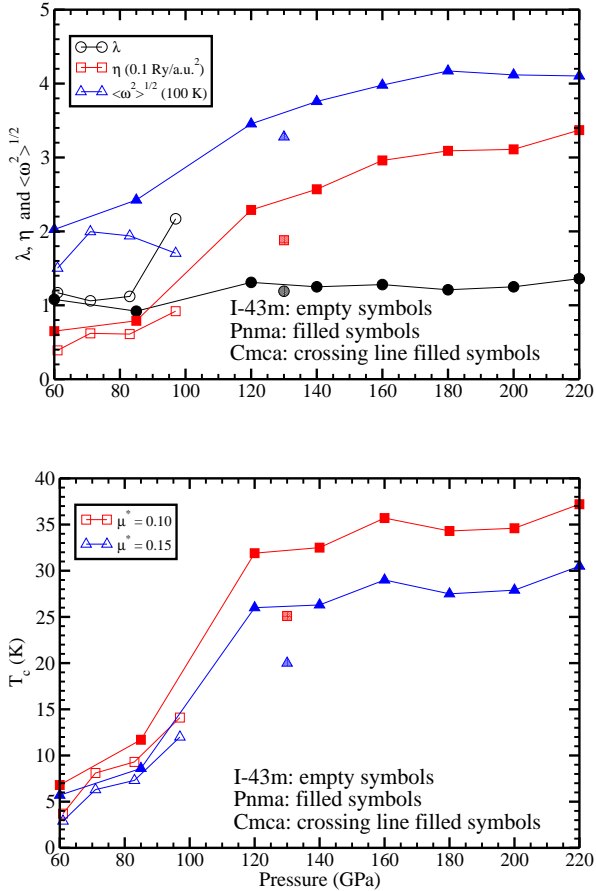

{\resizebox{7.8cm}{5.6cm}{\includegraphics{Fig6a.eps}}}
\vskip 6mm
{\resizebox{7.8cm}{5.6cm}{\includegraphics{Fig6b.eps}}}
\caption{Upper panel: Calculated electron-phonon coupling constant 
$\lambda$, $\eta$ and T$_C$ of 
Ca in I${\bar 4}$3m (empty symbols), Pnma (filled symbols) 
and Cmca (crossing-line filled symbols) structures at a few pressures. 
Lower panel: T$_c$ calculated from the Allen-Dynes equation, showing the
dependence on the Coulomb pseudopotential for which two values, $\mu^*$=0.10
and 0.15 have been taken.
   }
\label{T$_c$s}
\end{figure}

Figure \ref{T$_c$s} shows the calculated $\lambda$, 
$\eta = M_{Ca} <\omega^2>\lambda$, and rms frequency $<\omega^2>^{1/2}$ versus
pressure for a few structures and pressures.  The calculated values of T$_c$ are shown
in the lower panel, using two values of Coulomb pseudopotential $\mu^*$=0.10 and 0.15
that bracket the commonly used values and therefore
give an indication of the uncertainty due to the lack of knowledge of the
value of $\mu^*$ and its pressure dependence.  Results are provided for
Ca in I${\bar 4}$3m, Pnma and Cmca structures at a few pressures up to 220 GPa.
In elemental metals and in compounds where coupling is dominated by one atom type,
$\eta$ has often been useful in characterizing contributions to T$_c$.\cite{wep1982}
$\eta$ increases with pressure monotonically by a factor of more than 5 from 60 GPa to 220 GPa.
The coupling constant $\lambda$ increases modestly up to 120 GPa then remains nearly constant
at $\lambda$ = 1.2-1.4.
As pointed out elsewhere,\cite{Li2} a dense zone sampling is needed to calculate $\lambda$ accurately,
so any small variation is probably not significant.
The increase in $\eta$ beyond 120 GPa correlates well with the
lattice stiffening (increase in $<\omega^2>$) in this pressure range. 

The trend of the resulting T$_c$ generally follows, but seems to overestimate somewhat,
the experimental values\cite{Ca-exp}.
For Cmca structure at about 130 GPa,
the calculated EPC strength is $\lambda = 1.2$ and T$_c$ = 20-25K (for the two values of $\mu^*$)
in very satisfactory agreement with the observed values of T$_c$ in this pressure range.
For Pnma structure, T$_c$ increases rapidly in the 80-120 GPa region. At pressures above 120 GPa
up to the maximum 220 GPa that we considered,
the EPC constant $\lambda$ is $\sim$ 1.2-1.4 and 
the calculated T$_c$ increases modestly from 25-30K at 120 GPa to 30-35K at 220 GPa.  Neither
the structure dependence nor the pressure dependence seems very important:  
the strong coupling
and high T$_c$ is more the rule than the exception.
Ca at high pressure may be an excellent superconductor regardless of its structure.

\vskip 2mm

\section{Summary}
Calculations of enthalpy versus pressure for five crystalline
phases of Ca (simple cubic and four distortions from it) indicate 
quasi-degeneracy, with enthalpy differences small enough that one might 
expect a locally disordered, highly anharmonic, fluctuating structure at room
temperature.  
Over most of the 30-150 GPa range, we find at least three crystal
phases whose enthalpies indicate they will compete strongly at room temperature.
The sc phase itself is badly
unstable dynamically (at T=0), but the observed ``sc" diffraction pattern
can be understood as a locally noncrystalline, highly
anharmonic phase derived from a spatially inhomogeneous
and dynamically fluctuating combination of these structures, with
most of them being straightforward distortions from the sc structure.
Such a scenario seems to account qualitatively for 
the XRD observations of a ``sc''
structure.

At pressures below 100 GPa, 
the quasi-degenerate structures tend to have soft branches or occasionally
lattice instabilities, which are associated with strong electron-phonon
coupling. 
In the pressure range of 110 to 130 GPa
three phases (P4$_3$2$_1$2, Cmca and Pnma) again become quasi-degenerate,
and again it seems likely there will be spacial and temporal fluctuations
between the structures. 
Of course other structures may come into play as well; 
Arapan, Mao, and Ahuja\cite{Arapan}
have proposed that the Pnma structure competes with an incommensurate
structure at high pressure.

As our other main result, we find that linear response calculation of the EPC strength
and superconducting T$_c$ accounts for its impressive
superconductivity in the high pressure regime 
and accounts in a broad sense for the strong increase of T$_c$ in the ``sc" phase.
At higher pressure beyond the current experimental limit (i.e., 161 GPa), T$_c$
still lies in the 20-30 K range for some phases that we have studied.
In fact strong electron-phonon coupling seems to be present in several phases
across a substantial high pressure range, although we have no simple picture
why such strong coupling should arise.  (The strong coupling in Li and Y  likewise
has no simply physical explanation.\cite{Li-th2,Y-th1})
These results may resolve some of the perplexing
questions on the structure and record high T$_c$ for an element, and should help
in obtaining a more complete understanding of the rich phenomena that arise in
simple metals at high pressure.

After submission of our manuscript, we became aware of a study by
Yao {\it et al.}\cite{Yao-PRL}. They performed structural studies of calcium
in the range 34-78 GPa using metadynamics and genetic algorithm
methods.  Their methods and results are complementary to ours,
with each approach providing its own insights.  
Connections of their work to ours is evident; for example, the I4$_1$/amd structure
they focused on is slightly distorted from simple cubic, as are the
structures that we study.  
Since its enthalpy is within 20 meV/Ca of the Pnma structure across this pressure range, 
their result is consistent with our explanation of the observation of the simple
cubic diffraction pattern at room temperature.  
Their linear response calculations of electron-phonon coupling and the resulting
T$_c$ are also consistent with the more extensive results that we
present.

\section{Acknowledgments}
This work was supported by DOE through the Scientific Discovery through
Advanced Computing program (SciDAC grant DE-FC02-06ER25794), and by DOE
grant DE-FG02-04ER46111 with important interaction from the Computational
Materials Science Network.  One of us (F.G.) acknowledges support from
NSF through grant OCI PetaApps 0749217.

\end{document}